\documentclass{aa} % for the letters 
\usepackage{natbib}
\usepackage{graphicx}
\usepackage{color}
\usepackage{txfonts}
\usepackage{hyperref}
\begin{document} 
\label{firstpage} 
    \title{Ethylene oxide and Acetaldehyde in hot cores}
    \titlerunning{Ethylene oxide and acetaldehyde in hot cores}
    \authorrunning{Occhiogrosso et al.}
    \author{A. Occhiogrosso\inst{1, 2}
           \and
           A. Vasyunin\inst{3, 4}
           \and
           E. Herbst\inst{3}
           \and
           S. Viti\inst{1}
           \and
           M. D. Ward\inst{5}
           \and
           S. D. Price\inst{6}
           \and 
           W. A. Brown\inst{7}
           }
    \institute{Department of Physics and Astronomy, University College London (UCL), 
              Gower Street, London, WC1E 6BT, UK \\
 %             \email:ao@star.ucl.ac.uk
           \and
              Visiting Research Fellow, Astrophysics Research Centre, Queen's University Belfast,
              University Road, Belfast, BT7 1NN, UK\\ 
           \and
              Department of Chemistry, University of Virginia
              McCormick Road,  Charlottesville, VA22904-4319, USA\\
           \and
              Visiting Scientist, Ural Federal University, Ekaterinburg, Russia\\
           \and
              Lawrence Berkeley National Laboratories 
              1 Cyclotron Road, Berkeley, CA94720, USA\\
           \and
              Department of Chemistry, University College London (UCL), 
              Gordon Street, London, WC1H 0AJ, UK\\
           \and
              Department of Chemistry, University of Sussex, 
              Falmer, Brighton, BN1 9QJ, UK \\
             }
    \date{Received ; accepted}
%\maketitle
\abstract
% 5 {} token are mandatory 
  % context heading (optional)
  % {} leave it empty if necessary  
   {Ethylene oxide (c-C$_{2}$H$_{4}$O), and its isomer acetaldehyde (CH$_{3}$CHO), are important complex organic molecules because of their potential role in the formation of amino acids. The discovery of ethylene oxide in hot cores suggests the presence of ring-shaped molecules with more than 3 carbon atoms such as furan (c-C$_{4}$H$_{4}$O), to which ribose, the sugar found in DNA, is closely related.}
  % aims heading (mandatory)
   {Despite the fact that acetaldehyde is ubiquitous in the interstellar medium, ethylene oxide has not yet been detected in cold sources. We aim to understand the chemistry of the formation and loss of ethylene oxide in hot and cold interstellar objects (i) by including in a revised gas-grain network some recent experimental results on grain surfaces and (ii) by comparison with the chemical behaviour of its isomer, acetaldehyde.}
  % methods heading (mandatory)
   {We introduce a complete chemical network for ethylene oxide using a revised gas-grain chemical model. We test the code for the case of a hot core. The model allows us to predict the gaseous and solid ethylene oxide abundances during a cooling-down phase prior to star formation and during the subsequent warm-up phase. We can therefore predict at what temperatures ethylene oxide forms on grain surfaces and at what temperature it starts to desorb into the gas phase.}
  % results heading (mandatory)
   {The model reproduces the observed gaseous abundances of ethylene oxide and acetaldehyde towards  high-mass star-forming regions. In addition, our results show that ethylene oxide may be present in outer and cooler regions of hot cores where its isomer has already been detected. Our new results are compared with previous results, which focused on the formation of ethylene oxide only.}
  % conclusions heading (optional), leave it empty if necessary 
   {Despite their different chemical structures, the chemistry of ethylene oxide is coupled to that of acetaldehyde, suggesting that acetaldehyde may be used as a tracer for ethylene oxide towards cold cores.}
\maketitle

\section{Introduction}
Heterocycles are complex organic molecules containing heavier elements, in addition to carbon, in their ring structures. Among this class of compounds, O-bearing species are particularly important in the study of the interstellar medium (ISM) because of their links to the presence of life. Ribose, in particular, is one of the molecules associated with the  molecular structure of DNA. Considering simpler heterocyclic molecules than ribose, we arrive at ethylene oxide, also called oxirane (c-C$_{2}$H$_{4}$O). Ethylene oxide is the smallest cyclic species containing oxygen, where it is bonded to two carbon atoms. Acetaldehyde (CH$_{3}$CHO), a non-cyclic isomer of ethylene oxide, has an important role as an evolutionary tracer of different astronomical objects \citep{Herbst09}. c-C$_{2}$H$_{4}$O was first detected in the ISM by \citet{Gottlieb73} and has been subsequently observed towards different astrochemical objects; particularly important is its discovery in regions of star formation by \citet{Johansson84} and by \citet{Turner91}. 

The formation of ethylene oxide in high-mass star-forming regions has been extensively investigated. It was first detected by \citet{Dickens97} in the galactic centre source Sgr B2(N) and its column densities were found to be in the 10$^{13}$--10$^{14}$ cm$^{-2}$ range \citep{Ikeda01}. Unlike ethylene oxide, acetaldehyde has been detected in the cold gas of the dark cloud TMC-1 \citep{Matthews85}, in quiescent regions \citep{Minh00}, and towards hot cores \citep{Nummelin98}. In all regions, acetaldehyde seems to be spatially correlated to CH$_{3}$OH. Their abundances vary within three orders of magnitude in the Galactic Center \citep{Requena08}; in particular, the ratio of CH$_{3}$OH-to-CH$_{3}$CHO ranges from $\sim$ 10 in cold dark clouds up to 100 in hot cores. However, deviations from these values are possible \citep{Bisschop07, Bisschop08}; in particular, \citet{Oberg11} found a ratio much greater than 100 towards the protostar SMM1 within the Serpens core. The ratio between the fractional abundance of acetaldehyde and ethylene oxide lies in the range 1--9 towards several high-mass star-forming regions \citep{Nummelin98, Ikeda01}. There are no observations of vinyl alcohol (CH$_{2}$CHOH), the second structural isomer of ethylene oxide, in star-forming regions with the exception of its detection towards the Galactic Centre by \citet{Turner01}.
Even in star-forming regions, ethylene oxide and acetaldehyde both have low rotational temperatures between 20 and 40 K  \citep{Nummelin98}. In these regions, the emission from acetaldehyde originates mainly in cool envelopes, where the rotational temperature may not be that much below the kinetic temperature, which may also be the case for ethylene oxide. However, recent observations by \citet{Belloche13} towards the Galactic Center found a rotational temperature of 100 K associated with both ethylene oxide and acetaldehyde.

Laboratory experiments have been performed with the aim of determining the potential mechanisms of production for ethylene oxide and acetaldehyde under interstellar conditions. Using electron irradiation of mixtures of CO and CH$_{4}$ as well as  CO$_{2}$ with C$_{2}$H$_{4}$ ices, \citet{Bennett05} looked at the formation of acetaldehyde, ethylene oxide and vinyl alcohol in the solid state. They found that acetaldehyde forms with both mixtures, while ethylene oxide and vinyl alcohol can only be produced from irradiation of carbon dioxide and ethylene as reactants. In addition, \citet{Ward11} recently studied the solid state formation of ethylene oxide in a single reaction involving atomic oxygen and ethylene, which were accreted onto a solid substrate. 

These recent laboratory data suggest potential routes for the formation of ethylene oxide on grain surfaces started by an association reaction between atomic oxygen and ethylene. In the gas phase ethylene reacts with O atoms by readily breaking its C--C double bond \citep{Occhiogrosso13} in order to produce CO, instead of involving addition reactions in which O is attached across the double C--C bond to form the cyclic species. The feasibility of a new solid state route for ethylene oxide production under hot core conditions has recently been studied by \citet{Occhiogrosso12}. In particular, the authors were able to match the observed abundances of gaseous ethylene oxide in regions of star-formation where temperatures were high enough for the desorption of water ices to occur, and for ethylene oxide to sublime via co-desorption with water at temperatures at and above 100 K. Note that in light of the low rotational excitation temperature of ethylene oxide \citep{Nummelin98} the radiation may be emitted in the outer envelope, which would imply that ethylene oxide, like acetaldehyde, desorbs at lower temperatures in the collapse and heat up towards the protostellar core, below the 100 K assumed by \citet{Occhiogrosso12}.

In this paper, we extend our study of ethylene oxide formation by including a more complex chemistry for this molecule. We also insert a complete reaction network for the formation and destruction of one of its isomers, acetaldehyde. Given the lack of observational data on vinyl alcohol in star-forming regions, we do not include this isomer in our calculations. 
 
For the purpose of the present study, we employ an alternative code labelled MONACO \citep{Vasyunin13} to our standard UCL\_CHEM code and use a different approach for the treatment of the experimental data for the formation of c-C$_{2}$H$_{4}$O from atomic oxygen and ethylene \citep{Ward11}. We do not attempt to include the high-energy mechanisms of \citet{Bennett05} since neither code is equipped to model these processes effectively.
In Section 2, we describe the details of the MONACO chemical model and we dedicate a subsection (2.1) to the computational treatment of the laboratory data. In Section 3, we present the results from the MONACO network and code, and we report a comparison with previous theoretical and observational work in subsection 3.1. Our conclusions are given in Section 4.

\begin{table*}[ht]
 \begin{center}
  \caption{Non-zero initial abundances of species with respect to {\it n$_{H}$} used in the MONACO chemical model, based on \citet{Wakelam08}.}
  \begin{tabular}{ccccccc}
  \hline
H & H$_{2}$ & He & N & O & C$^{+}$ & S$^{+}$ \\
  \hline 
1.0$\times$10$^{-03}$ & 4.9$\times$10$^{-01}$ & 9.0$\times$10$^{-02}$ & 7.6$\times$10$^{-05}$ & 2.6$\times$10$^{-04}$ & 1.2$\times$10$^{-04}$ & 8.0$\times$10$^{-08}$ \\
  \hline
Si$^{+}$ & Fe$^{+}$ & Na$^{+}$ & Mg$^{+}$ & Cl$^{+}$ & P$^{+}$ & F$^{+}$\\
  \hline
8.0$\times$10$^{-09}$ & 3.0$\times$10$^{-09}$ & 2.0$\times$10$^{-09}$ & 7.0$\times$10$^{-09}$ & 1.0$\times$10$^{-09}$ & 2.0$\times$10$^{-10}$ & 6.7$\times$10$^{-09}$ \\
  \hline
  \end{tabular}
 \end{center}
\end{table*}

\section{Methodology}

We use a modified version of the MONACO gas-grain chemical network/model, which was first implemented by \citet{Vasyunin09} and then extended in order to take some account of the internal monolayers below the surface \citep{Vasyunin13}. Although the latest version uses a Monte Carlo approach, the version used here is based on the rate equation method proposed by \citet{Hasegawa92} and applied to the treatment of the diffusion mechanism, treating the ice as a bulk system. The diffusion describes the mobility of species in the mantle and on its surface after they have been adsorbed. Specifically, molecules can jump between two adjacent sites either due to thermal energy (thermal hopping) or because of tunnelling effects. The theory behind these two mechanisms is explained in more detail in Subsection 2.1. In the MONACO model, molecular hydrogen, helium and ions are not allowed to stick on the grain surface, unlike all the other species, which accrete to it with a sticking coefficient of unity. 

The code is used to treat the chemistry that occurs during the formation of a dense and warm core, the so-called {\it hot core}. The term {\it hot core} refers to material collapsing into a protostar and heating up over time. The outer gas and dust are cool and the inner gas and dust are heated up as they proceed inwards. The so called {\it cold phase} during the hot core formation starts from a cold (20 K) and relatively diffuse medium ({\it n$_{H}$} = 3$\times$10$^{3}$ cm$^{-3}$) where all the species (with the exception of hydrogen) are in an atomic form. The non-zero initial fractional abundances with respect to the total number of hydrogen nuclei {\it n$_{H}$}  are taken from \citet{Wakelam08} and are reported in Table 1. The species, with the exception of hydrogen and helium, have  initial abundances significantly below those observed in diffuse interstellar clouds, with some metallic elements two orders of magnitude lower than the solar elemental abundances. These abundances are often referred to as ``low-elemental abundances'' \citep{Vasyunin13} and are often used in order to take into account additional elemental depletion on grains in cold interstellar cores. 

Instead of assuming a constant density during the cold pre-stellar-core stage, or assuming an isothermal collapse during this stage, we describe the initial stage as one in which the temperature decreases from 20 K to 10 K in order to account for the fact that the efficiency of radiative heating of grains drops as the visual extinction increases. At the same time, the density increases, mimicking a collapse to form a cold core in which much of the material is in icy mantles. Once the system reaches 10 K, the temperature starts to rise to 200 K in order to reproduce the warm-up phase leading to the formation of a hot core. As a consequence, molecules sublime from the grain mantle. During the initial stages of the warm-up phase, non-thermal desorption effects play an important role. They include desorption due to cosmic rays, UV-photodesorption \citet[treated according to the laboratory study by][]{Oberg07}, and reactive desorption in which the products of an exothermic surface reaction are immediately ejected into the gas phase \citep{Katz99, Garrod07}.  
The timescale adopted for the hot core formation is shown in Fig. 1 (left panel). In the right panel, we expand the fast increase in temperature occurring during the warm-up phase. The temperature profile is taken from \citet{Viti04} and adapted to account for the decrease in temperature from 20 K to 10 K during the collapse phase. The density increase during Phase I is also reported in Fig. 2. This simulation refers specifically to the case of a massive star of approximately 25 M$_{\odot}$. 

\begin{figure}
 \includegraphics[width=90mm]{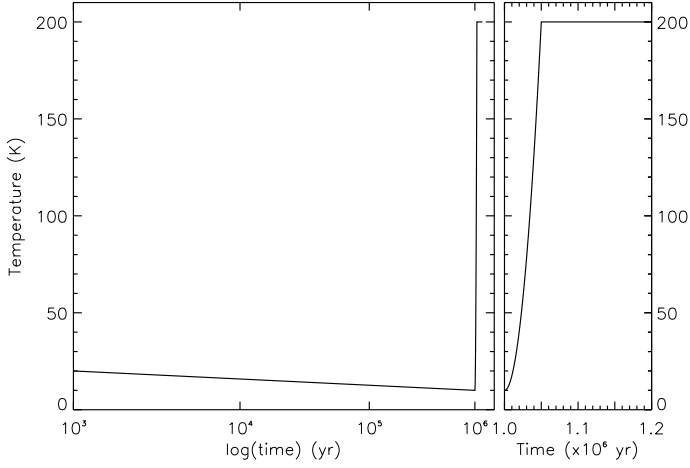}
 \begin{center}
 \caption{The temperature profile during the initial collapse phase and the subsequent warm-up phase leading to hot core formation plotted as a function of time. The panel on the right expands the final portion of the panel on the left.}
 \end{center}
\end{figure}
\begin{figure}
 \includegraphics[width=60mm, angle=270]{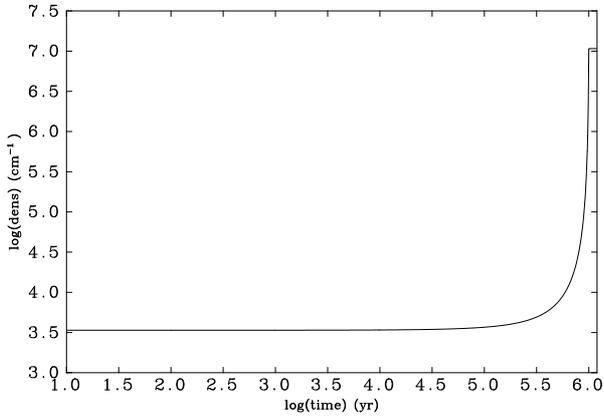}
 \begin{center}
 \caption{The density profile during the initial collapse phase and the subsequent warm-up phase leading to hot core formation plotted as a function of time.}
 \end{center}
\end{figure}
In the version of the MONACO code employed for the present study, no results from TPD experiments \citep{Collings04} are included in the treatment of the thermal-desorption, a treatment that here is taken from \citet{Tielens82}. The mathematical treatment of thermal desorption, or sublimation, is based on the first-order Polanyi-Wigner equation with respect to the abundance of a selected molecule:

\begin{equation}
k_{\rm subl} = \nu \times e^{-E_{\rm D}/T}
\end{equation}

\setlength\parindent{0pt} where the frequency {\it $\nu$}, often referred to as the trial frequency, contains a dependence on the molecular mass and {\it E$_{D}$} is the desorption energy for a selected species. The trial frequency for physisorbed H atoms is normally assumed to be $\sim$ 3$\times$10$^{12}$ s$^{-1}$ \citep{Herbst08}.
In the present study, we use a value of 2450 K \citep[appropriate for a water-ice surface,][]{Garrod06b} for the desorption energy for both ethylene oxide and acetaldehyde.
During the cold collapse stage, there is little to no thermal desorption (sublimation) for species heavier than hydrogen and helium.  During the warm-up stage, the icy mantles eventually totally sublime, but not before an active mantle chemistry leads to the production of complex molecules \citep{Vasyunin13}.

\begin{table*}
 \begin{center}
 \caption{Gas phase rate parameters for acetaldehyde formation and destruction taken from the KIDA database \citep{Wakelam12}.}
 \begin{tabular}{cccccrrr}
  \hline
R1 & R2 & P1 & P2 & P3 & $\alpha$ (cm$^{3}$s$^{-1}$) & $\beta$ & $\gamma$ (K)\\[0.05ex]
   \hline
\\[0.01ex]  
CH$_{3}$CHO & He$^{+}$ & He & CH$_{3}$ & HCO$^{+}$ & 1.40$\times$10$^{-09}$ & -0.50 & 0.00\\
CH$_{3}$CHO & He$^{+}$ & He & CH$_{3}$CO$^{+}$ & H & 1.40$\times$10$^{-09}$ & -0.50 & 0.00\\
CH$_{3}$CHO & H$^{+}$ & H$_{2}$ & CH$_{3}$CO$^{+}$ & & 5.00$\times$10$^{-09}$ & -0.50 & 0.00\\
CH$_{3}$CHO & O & CH$_{3}$CO & OH & & 1.79$\times$10$^{-11}$ & 0.00 & 1100.00\\
CH$_{3}$CHO & OH & CH$_{3}$CO & H$_{2}$O & & 1.42$\times$10$^{-11}$ & 0.00 & 0.00\\
CH$_{3}$CHO & CH$_{3}$O & CH$_{3}$OH & CH$_{3}$CO & & 8.30$\times$10$^{-15}$ & 0.00 & 0.00\\
CH$_{3}$CHO & H & CH$_{3}$CO & H${2}$ & & 6.64$\times$10$^{-11}$ & 0.00 & 2120.00\\
CH$_{3}$CHO & N & H$_{2}$ & HCN & HCO & 1.99$\times$10$^{-14}$ & 0.00 & 0.00\\
CH$_{3}$CHO & CH$_{3}$$^{+}$ & C$_{3}$H$_{6}$OH$^{+}$ & photon & & 5.70$\times$10$^{-11}$ & 0.66 & 0.00\\
CH$_{3}$CO & C$_{2}$H$_{6}$ & CH$_{3}$CHO & C$_{2}$H$_{5}$ & & 1.95$\times$10$^{-13}$ & 2.75 & 8830.00\\
CH$_{3}$CO & H$_{2}$ & CH$_{3}$CHO & H & & 2.21$\times$10$^{-13}$ & 1.82 & 8870.00\\
CH$_{3}$CO & H$_{2}$CO & CH$_{3}$CHO & HCO & & 3.01$\times$10$^{-13}$ & 0.00 & 6510.00\\
CH$_{3}$CO & CH$_{4}$ & CH$_{3}$CHO & CH$_{3}$ & & 4.92$\times$10$^{-14}$ & 2.88 & 10800.00\\
CH$_{3}$CO & HCO & CH$_{3}$CHO & CO & & 1.50$\times$10$^{-11}$ & 0.00 & 0.00\\
   \hline
 \end{tabular}
 \end{center}
\centering
\tiny R$_{1}$ and R$_{2}$ indicates the reactants and P$_{1}$, P$_{2}$, P$_{3}$ are the different products of reaction; $\alpha$, $\beta$, and $\gamma$ are parameters for the rate coefficients (see Equation (2)).
\end{table*}

The chemical network of the MONACO model is based on the OSU gas-grain database (http://www.physics.ohio-state.edu/$\sim$eric/research.html) and we introduce the following changes for the purpose of the present study: 
\begin{enumerate}
\item We insert a complete gas phase reaction network for ethylene oxide based on Table 3 in \citet{Occhiogrosso12}.
\item We include the grain surface route for the formation of ethylene oxide investigated by \citet{Ward11} in which oxygen atoms associate with ethylene on growing icy mantles covering a substrate of graphite (HOPG). The manner in which we utilise and adapt the experimental results of \citet{Ward11} is explained in detail in Subsection 2.1. Acetaldehyde is produced either by several gas phase routes, or via an association reaction in the ice mantle. 
\item We update some of the gas phase reaction rate coefficients involving acetaldehyde using values from the KIDA database \citep{Wakelam12} and we report them in Table 2.
\end{enumerate}
The rate parameters can be inserted in the following formula (also known as the Kooij formula) in order to evaluate the rate coefficients in units of cm$^{3}$ s$^{-1}$:

\begin{equation}
 k(T) = \alpha(T/300)^{\beta}e^{-\gamma/T}, 
\end{equation}
where $\alpha$, $\beta$ and $\gamma$ are the parameters characterizing the variation of the rate coefficients with temperature.
Reaction rates are calculated by solutions of coupled first-order differential equations with respect to the evolution time of the core, depending on the reactant concentrations and on the rate coefficients. 
The outputs from the code are fractional abundances $f$(X) with respect to the total density of hydrogen nuclei, which can be easily converted into column densities using the following equation:

\begin{equation}
N(X) = f(X)\times A_{\rm v} \times N(H + 2(H_{2})),
\end{equation} 
where {\it A$_{v}$} is the visual extinction and {\it N(H + 2(H$_{2}$))} is equal to 1.6$\times$10$^{21}$ cm$^{-2}$, the hydrogen column density corresponding to one mag.

\subsection{Computational treatment of experimental data}

In their study of ethylene oxide formation, \citet{Ward11} reported a value of 1.2$\times$10$^{-19}$ cm$^{2}$ s$^{-1}$ for the bimolecular reaction rate coefficient of the association of O and C$_{2}$H$_{4}$ at 20 K. Via the temperature dependence of the rate coefficient, they also determined a total energy barrier of 190 K. The latter quantity represents the contribution to the system of the activation energy as well as the diffusion barrier, E$_{b}$, if we assume the reaction to occur via diffusion. 
The  rate coefficient for the formation of ethylene oxide can then be expressed as
\begin{equation}
k = a \times e^{-190/T} 
\end{equation}
from  which it follows by substitution of the value at 20 K that {\it a} = 1.6$\times$10$^{-15}$ cm$^{2}$s$^{-1}$. In order to obtain the rate coefficient in units of s$^{-1}$, appropriate if the amounts of  reactants are expressed as numbers   rather than real concentrations, we multiply {\it a} by the site density {\it N} = 1.5$\times$10$^{15}$ cm$^{-2}$ \citep{Hasegawa92} to determine a value of $a$ equal to 2.4 s$^{-1}$ .

If we assume that the reaction proceeds via diffusion, a mechanism known in surface science as the Langmuir-Hinshelwood mechanism, we can also make the approximation that at low temperatures, the diffusion is dominated by light atomic species, in this case oxygen.  
As reported by \citet{Herbst85}, if we neglect the effect due to quantum tunnelling, the hopping rate coefficient can then be expressed by the following equation:
\begin{equation}
k (s ^{-1} ) =  \nu \times e^{-E_{\rm b}/T}
\end{equation}

\setlength\parindent{0pt} where {\it $\nu$} is the vibrational frequency of the lighter adsorbate \citep[typically 10$^{12-13}$ s$^{-1}$,][]{Herbst08} in a potential well, {\it E$_{b}$} is the diffusion barrier (in units of K) between two adjacent binding sites, and {\it T} is the temperature in K. For a system in which there is also chemical activation energy, the diffusion barrier must be replaced by a total energy, which we label {\it E$_{\rm tot}$}.

The first-order rate coefficients for diffusion are defined as the hopping rates over a number of reactive sites equivalent to the total number of sites on the dust particles. The concentration of surface species is expressed in units of numbers per grain. Starting with a trial frequency of 3$\times$10$^{12}$ s$^{-1}$ and a total number of 10$^{6}$ sites for a grain radius of 0.1 $\mu$m, we obtain a reduced frequency of  3$\times$10$^{6}$ s$^{-1}$. 
To obtain the theoretical total barrier {\it X} in the diffusive approach,  we equate the laboratory formula for the rate coefficient  with the theoretical formula:
\begin{equation}
2.4 \times e^{-190/T} = 3\times10^{6} e^{-E_{\rm tot}/T}.
\end{equation}
Note that according to eq. (6) the energy barrier for the diffusion model shows a temperature dependence. Following the approach by \citet{Fuchs09}, who fitted their experimental data to a variable activation barrier, we obtain
\begin{equation}
E_{\rm tot}(T) = 190+14.22\times T.
\end{equation} 
Whether or not this temperature variation makes sense in this case is not a question we can answer here. Whether we use the modified experimental rate coefficient in units of s$^{-1}$ or the diffusion model value makes no difference to the analysis here since the rate coefficients for each approach are obviously equal to each other for all temperatures.  

We also point out that the experimental expression (or its diffusion model equivalent)  used is only valid over a temperature range of 10--90 K. Despite the fact that we run our models up to 200 K, any deviation from Arrhenius behaviour in the fit to the experimental data will not affect our results as ethylene oxide and acetaldehyde both desorb at temperatures much lower than 90 K ($\sim$65 K, see the following section for more details).

\section{Modelling}
In the present study we adopt two different sets of physical conditions. First, we consider a one-phase model that runs at a single temperature and density. Secondly, as mentioned previously, we consider a more complex two-phase model.

We start by inserting a complete chemical network for ethylene oxide into the MONACO network. We include the gas phase channels reported in \citet{Occhiogrosso12}, where the rate for each channel is calculated by multiplying the global rate taken from the OSU database by the braching ratios. In the gas phase reaction network, c-C$_{2}$H$_{4}$O possesses both formation and destruction channels. On the other hand, the reaction between O and C$_{2}$H$_{4}$ is the only route forming ethylene oxide on the grain surface. We complete the reaction scheme by introducing other desorption mechanisms for solid ethylene oxide.

\subsection{Single temperature models}
Single temperature models have been run in order to predict the ethylene oxide fractional abundances at different specific temperatures and densities. The different physical parameters adopted for these preliminary models are listed in Table 3 along with the ethylene oxide fractional abundances in the gas phase as well as in the icy mantle (the $m$ before the molecular formula stands for ``mantle''). 
We assume for simplicity that the dust and gas temperatures are the same, an assumption normally made for dense clouds although not rigorously confirmed. We utilise different temperatures as well as densities in order to test the sensitivity of the molecular abundances to the changes in these parameters. In order to emphasise the effects due to non-thermal desorption (N-td), we run models with these processes ON and OFF. 

\begin{table}
 \begin{center}
 \caption{Results for gaseous and solid ethylene oxide fractional abundances at 1.2$\times$10$^{6}$ years in different one-temperature chemical models.}
 \begin{tabular}{llllll}
  \hline
 & n$_{H}$/cm$^{-3}$ & T/K & N-td & $f$(c-C$_{2}$H$_{4}$O) & $f$($m$c-C$_{2}$H$_{4}$O) \\[0.05ex]
   \hline
\\[0.01ex]  
M1 & 1x10$^{4}$ & 10 & ON & 5.7$\times$10$^{-14}$ & 6.8$\times$10$^{-18}$\\
M2 & 1x10$^{4}$ & 15 & ON & 3.7$\times$10$^{-14}$ & 5.5$\times$10$^{-18}$\\
M3 & 1x10$^{4}$ & 20 & ON & 6.3$\times$10$^{-15}$ & 1.1$\times$10$^{-18}$\\
M4 & 1x10$^{4}$ & 20 & OFF & 1.9$\times$10$^{-17}$ & 1.1$\times$10$^{-12}$\\
M5 & 1x10$^{4}$ & 50 & ON & 4.6$\times$10$^{-15}$ & 1.2$\times$10$^{-18}$\\
M6 & 1x10$^{4}$ & 50 & OFF & 2.0$\times$10$^{-21}$ & 6.4$\times$10$^{-25}$\\
M7 & 1x10$^{7}$ & 200 & ON & 3.3$\times$10$^{-16}$ & 2.1$\times$10$^{-32}$\\
M8 & 1x10$^{7}$ & 200 & OFF & 3.3$\times$10$^{-16}$ & 2.1$\times$10$^{-32}$\\
M9 & 1x10$^{8}$ & 300 & ON & 4.3$\times$10$^{-18}$ & 5.8$\times$10$^{-35}$\\
M10 & 1x10$^{8}$ & 300 & OFF & 4.3$\times$10$^{-18}$ & 5.8$\times$10$^{-35}$\\
   \hline
\\[0.01ex]
 \end{tabular}
\end{center}
%\centering
\tiny Starting from the left, the first column represents the gas density; T is the temperature of the source; the third column shows whether the model accounts (ON) or not (OFF) for the effects due to non-thermal desorption; $f$(c-C$_{2}$H$_{4}$O) and $f$($m$c-C$_{2}$H$_{4}$O) are the fractional abundances of ethylene oxide relative to hydrogen nuclei in the gas-phase and on the grain surface, respectively. The {\it m} before the molecular formul$\ae$ stands for {\it mantle}.
\end{table}

Our results are summarised in Table 3, which reports the fractional abundance of ethylene oxide in the gas phase, $f$(c-C$_{2}$H$_{4}$O), as well as on the grain surface, $f$($m$c-C$_{2}$H$_{4}$O), at the end of the simulation. They show that, regardless of the different physical parameters chosen and assuming that all the solid ethylene oxide sublimes to the gas phase, the models are unable to produce gaseous ethylene oxide above the detection threshold ($\sim$10$^{-12}$). A general conclusion is that non-thermal desorption is critically important at low temperatures, where thermal desorption is not efficient. This is unequivocal if we look at the fractional abundances of ethylene oxide obtained as output from the models at higher temperatures with and without the influence due to non-thermal desorption effects: in both cases the amount of ethylene oxide is still the same. Some effects mainly due to reactive desorption can be highlighted at 50 K (see M5 and M6), but the major consequences due to non-thermal desorption are manifest at 20 K (see M3 and M4) where ethylene oxide is not efficiently produced on the grain surfaces when the mechanism is operating.

\begin{figure*}
 \begin{center}
\includegraphics[width=90mm, angle=90]{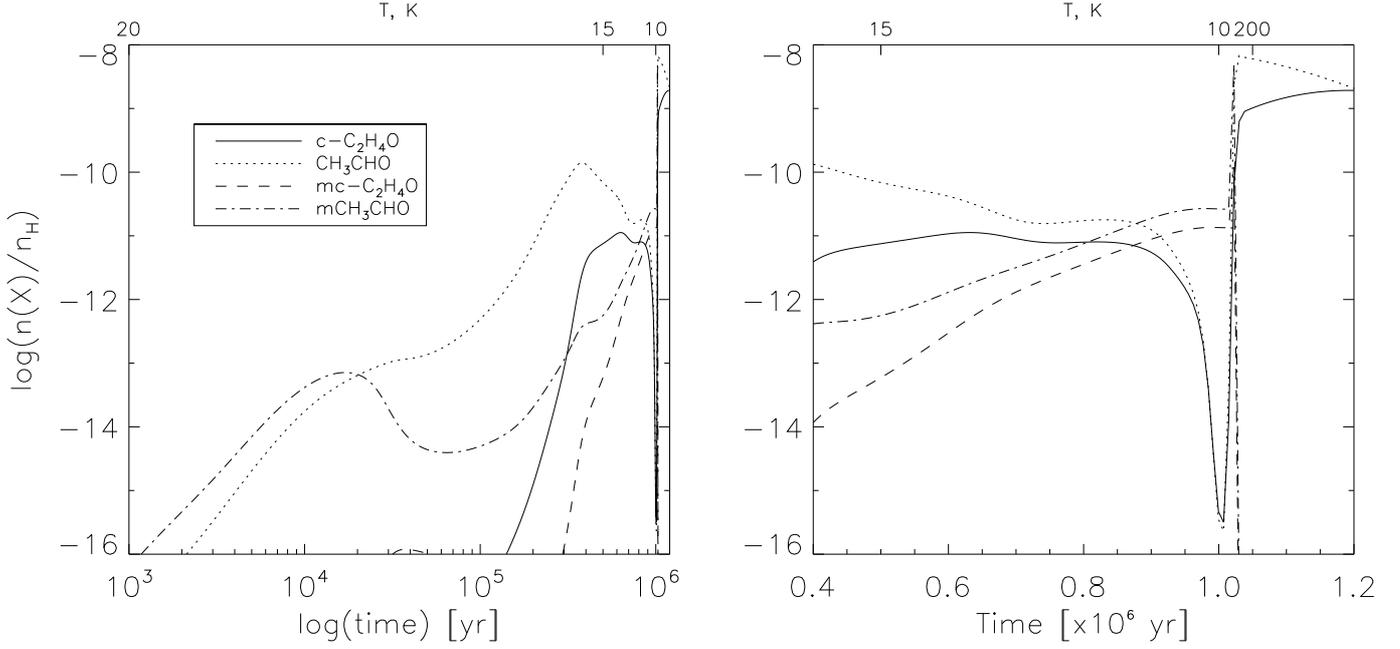}
\caption{Fractional abundances of gaseous and solid ethylene oxide and acetaldehyde as a function of time during two phases, over 10$^{6}$ yr. The panel on the right expands the final portion of the panel on the left. The designation $m$ refers to the grain mantle.}
 \end{center}
\end{figure*}

\subsection{Two-phase models}
\indent Single temperature models are advantageous in terms of computational time but the chemical scenario may not be realistic because reactants can clearly desorb before they interact to produce more complex species, depending upon the temperature chosen. Moreover, during star-formation we assume an increase in temperature when the pre-stellar core starts to irradiate. Hence, we ran models where the temperature eventually increases as a function of time. This helps to produce complex molecules in the gas phase because at temperatures usually T $\ge$ 25 K, molecular reactants on the grain start to diffuse appreciably and react after photo-activation to form complex molecules. The complex molecules formed on the grains then either start to evaporate as the temperature continues to rise or they desorb non-thermally \citep{Garrod06b, Garrod08}.
As previously mentioned, the warm-up phase model runs in two stages. First, we follow the trend for the free-fall collapse during which we observe an increase in density from 3$\times$10$^{3}$ cm$^{-3}$ (in a diffuse/translucent medium) to 10$^{7}$ cm$^{-3}$ (which is the typical density of a hot core) and a decrease in temperature from 20 K to 10 K  to account for the fact that the efficiency of radiative heating of grains drops as the visual extinction increases \citep{Vasyunin13}. The visual extinction is initially set equal to 2 mags and by the end of Phase I it rises to $\sim$ 430 mag. The simulation continues until the final density is reached, which takes over 10$^{6}$ years. The accretion efficiency is controlled by the sticking probability of gaseous atoms and molecules, which is assumed to be unity.
During Phase II, the density is kept constant, while the temperature goes from 10 K up to 200 K over 5$\times$10$^{4}$ years. By the end of this second phase a hot core is formed. We consider the process of formation for a massive star of 25 M$_{\odot}$. Note that lower masses lead to different desorption times as a consequence of the different temperature profiles. Specifically, assuming a monotonic increase of the gas temperature with the luminosity of a star (and therefore the age), \citet{Viti04} found that the sublimation of molecules in low-mass stars is shifted towards higher temperatures due to the slowing down of the heating rate.

\section{Results}
Figure 3 shows the calculated abundances of gaseous and solid ethylene oxide and acetaldehyde from the two-stage model as a function of time. The solid and dashed lines represent the abundances of gaseous and icy ethylene oxide, while the dotted and dashed-dotted lines represent the abundances of gaseous and solid state acetaldehyde. 
The panel on the left reports results from the collapse and warm-up phases, while that on the right expands the results from the warm-up phase, where desorption eventually leads to large abundances of gaseous ethylene oxide and acetaldehyde.

\subsection{Analysis of the ethylene oxide abundance profile}
In the gas phase, the dominant mechanism of formation of ethylene oxide is the dissociative recombination of ciclic C$_{2}$H$_{5}$O$^{+}$: 
\begin{equation}
\rm{e^{-} + C_{2}H_{5}O^{+} \rightarrow c-C_{2}H_{4}O + H},
\end{equation} 

while the major loss pathway is the reaction between ethylene oxide and C$^{+}$ which produces C$_{2}$H$_{3}$O$^{+}$ and CH or C$_{2}$H$_{4}$O$^{+}$ and C. Photodissociation reactions also occur but are important only at the very early stage of star formation (during the collapse phase) when the density and hence the visual extinction are low enough for the external UV to penetrate the cloud core. The photodissociation of ethylene oxide leads to the formation of CH$_{3}$ and HCO or CH$_{4}$ and CO. Finally, ethylene oxide is protonated by reactions with HCO$^{+}$, H$_{3}$$^{+}$ and H$_{3}$O$^{+}$. During the collapse phase, part of the gaseous ethylene oxide freezes onto the grain-surface; at this stage, only a small amount of ethylene oxide is returned into the gas-phase due to non-thermal desorption effects. 

Concerning the chemistry of solid ethylene oxide, while the initial diffuse medium is collapsing, ethylene oxide forms slowly on the grain surface via the reaction between oxygen atoms and ethylene. The latter species is in turn produced mainly by the hydrogenation of C$_{2}$H$_{2}$ and via the freeze-out of gaseous ethylene. Solid ethylene is also formed by the surface reactions between solid CH$_{3}$ and CH and between H and C$_{2}$H$_{3}$; moreover, CH$_{2}$ dimerizes to C$_{2}$H$_{4}$. The final fractional abundances of solid ethylene in all our models are about three orders of magnitude lower than those of methane on the grain surface.
The abundance of solid atomic oxygen required to form a detectable amount of ethylene oxide has been extensively investigated by \citet{Occhiogrosso12}, who found that the 1$\%$ of oxygen in the icy mantle would be enough to allow the reaction with ethylene and produce an observable amount of c-C$_{2}$H$_{4}$O. In the present paper, we therefore do not discuss this point any further. 
Solid ethylene oxide reaches a peak fractional abundance on the order of $\times$10$^{-11}$ just before 10$^{6}$ years.
When the warm-up phase starts, c-C$_{2}$H$_{4}$O sharply decreases on the grain surface because of changes in the physical conditions that make its production inefficient in the solid state and enable its gas phase formation (see left panel, Fig. 3). During Phase II, the desorption of c-C$_{2}$H$_{4}$O from the icy mantles contributes to an increase in its gas phase abundance; we can observe a sharp increment up to $\sim$2$\times$10$^{-9}$ in the abundance of gaseous ethylene oxide after just 10$^{6}$ years. According to \citet{Viti04}, at this stage the cloud temperature is still too low (about 65 K) for most of the icy molecules to desorb, with the exception of CO-like species \citep{Collings04} and simple hydrocarbons like methane. Despite the cloud being quite cold, results from MONACO model highlight the possibility for ethylene oxide to desorb thermally due to a low binding energy to the surface; moreover the rotational temperature for ethylene oxide has been estimated to be up to 40 K only unlike many other complex species \citep{Nummelin98}.  This value is low enough to be consistent with the early sublimation of solid ethylene oxide (compared to other molecules of similar complexity) reported in Fig. 3 (right panel). 

At high temperature, apart from desorption, following the solid phase O + C$_{2}$H$_{4}$ reaction, another important contribution to the high abundance of ethylene oxide in the gas phase comes from dissociative recombination (see Reaction 8).
Thanks to the combined contribution of desorption and dissociative recombination, ethylene oxide reaches a plateau in its gaseous abundance, which lasts until the end of our simulation (1.2$\times$10$^{6}$ years). This plateau is lower in abundance than the peak abundance and so this trend is consistent with a low rotational temperature.

\subsection{Analysis of the acetaldehyde abundance profile}
In the gas-phase, acetaldehyde is mainly produced by the reaction between O and C$_{2}$H$_{5}$ and is destroyed by reaction with C$^{+}$ to form C$_{2}$H$_{3}$O$^{+}$ and CH or C$_{2}$H$_{4}$O$^{+}$ and C.
As with the case of ethylene oxide, another gas phase contribution to acetaldehyde formation comes from dissociative recombination of C$_{2}$H$_{5}$O$^{+}$:
\begin{equation}
\rm{ e^{-} + C_{2}H_{5}O^{+} \rightarrow CH_{3}CHO + H}
\end{equation}
assuming that there is a 50$\%$ probability for the cyclic ion to form a non-cyclic structure upon recombination.

The ionic species (C$_{2}$H$_{5}$O$^{+}$) is in turn formed via several gas phase reactions including the one between CH$_{4}$ and H$_{2}$CO$^{+}$, which represents the most favourable reaction pathway. The ion is also formed by so-called ``loop'' reactions, in which neutral acetaldehyde is protonated by ions such as H$_{3}^{+}$ and H$_{3}$O$^{+}$: 

\begin{equation}
\rm{CH_{3}CHO + H_{3}^{+} \rightarrow C_{2}H_{5}O^{+} + H_{2}}
\end{equation}

\begin{equation}
\rm{CH_{3}CHO + H_{3}O^{+} \rightarrow C_{2}H_{5}O^{+} + H_{2}O}
\end{equation}

Such reactions can also be considered as destruction reactions of acetaldehyde as long as the protonated ion does not fully re-convert back to acetaldehyde.

On the grain-surface, when temperatures are $\sim$30 K or above, acetaldehyde is formed by the radical-radical association reaction between CH$_{3}$ and HCO. Moreover, part of the gaseous acetaldehyde freezes onto the grain-surface until the temperature is warm enough for thermal desorption. As in the case of ethylene oxide, non-thermal desorption mechanisms due to secondary cosmic-ray photons have only a negligible effect in desorbing acetaldehyde from the icy mantles. However, gas phase reaction pathways produce such a large amount of this molecule that the gaseous acetaldehyde abundance exceeds the amount of its isomer up to a factor of 3.5 at the peak value during the warm-up phase. This larger abundance of acetaldehyde compared to the one of ethylene oxide is in agreement with the ratio between the two isomers reported in the literature; in particular, this ratio ranges from 1 to 9 based on the study by \citet{Ikeda01}. 
 
Acetaldehyde (as well as ethylene oxide) shows a peak in its gaseous abundance at $\sim$ 6.8$\times$10$^{-9}$ immediately after 10$^{6}$ years when the warm-up phase has just started, which corresponds to a temperature of $\sim$65 K. Since we found a peak in the acetaldehyde abundance at 65 K, its rotational temperature is likely to be lower than this value, unless the density exceeds the critical one. Finally, the gas phase abundance of acetaldehyde drops to 1.9$\times$10$^{-9}$ by the end of our simulation due to the reaction with H$_{3}$O$^{+}$ to form C$_{2}$H$_{5}$O$^{+}$ and H$_{2}$O (see reaction 11); the ion reacts with electrons in turn to form ethylene oxide.

\subsection{Discussion}
In this section we compare our hot core results with those from the models developed in a previous study by \citet{Occhiogrosso12}, as well as with observations. 

\citet{Occhiogrosso12} used UCL\_CHEM, a gas-grain chemical model, to investigate the chemistry of ethylene oxide towards high-mass star-forming regions and they were able to reproduce its observed abundances in these environments.
\begin{table}
 \begin{center}
  \caption{Comparison between the fractional abundances of the two isomers obtained from MONACO at their peak value and those from the study by \citet{Occhiogrosso12}.}
  \begin{tabular}{@{}ccc}
  \hline
 & c-C$_{2}$H$_{4}$O & CH$_{3}$CHO\\
   \hline
MONACO & 1.9$\times$10$^{-9}$ & 6.8$\times$10$^{-9}$\\
$^{a}$UCL\_CHEM & 1.0$\times$10$^{-10}$ & - \\
   \hline
  \end{tabular}
 \end{center}
\centering
\tiny $^{a}$Results refer to the fractional abundances of ethylene oxide at 200 K and 1.2$\times$10$^{6}$ yrs.
\end{table} 

In Table 4 we therefore report the fractional abundances for the two isomers achieved with the MONACO model for a hot core after 10$^{6}$ yrs and those from the best model by \citet{Occhiogrosso12} for the case of a massive protostar of 25 M$_{\odot}$ at 200 K. At the final time, the fractional abundance of ethylene oxide obtained using the UCL\_CHEM model is lower by one order of magnitude compared with that obtained using MONACO. Note that the simulations by \citet{Occhiogrosso12} did not take into account the chemistry of acetaldehyde and this factor may affect the results.
Moreover, relatively high temperatures ($\sim$ 100 K) were needed in order to desorb ethylene oxide from the grain surface to the gas phase; hence, they did not predict the peak of the ethylene oxide abundance at around 65 K as found by running the MONACO model.

In light of these dissimilarities, we have revised our previous results \citep{Occhiogrosso12} by updating the UCL\_CHEM model with a complete network of reactions for acetaldehyde as shown in Table 2. We also include grain-chemistry for this species taken from the MONACO model. Although the two codes present important differences in the way they treat the physics and chemistry under ISM conditions, we minimised these divergences by using in UCL\_CHEM the same physical parameters as in the simulations performed with the MONACO code. In particular, we adopt the same initial abundances (see Table 1), the same temperature profile (see Figure 1) and the same values for the final density and temperature of the core.  
One of the key issues concerns the treatment of the thermal desorption of species occurring during the warm-up phase. In the study by \citet{Occhiogrosso12}, the authors adopted the classification of species developed by \citet{Viti04} and they set the thermal desorption of ethylene oxide from the grain surface to occur at $\sim$100 K via co-desorption with water ices. In order to investigate the presence of a peak of ethylene oxide at lower temperatures, we now assign ethylene oxide as intermediate between CO- and H$_{2}$O-like molecules in their desorption pattern \citep{Viti04}. This signifies that ethylene oxide will undergo mono-molecular desorption (which will depend on its binding energy of $\sim$2450 K), volcanic desorption (when the amorphous-to-crystalline H$_{2}$O-ice conversion occurs) and co-desorption with water ice. Based on the similarities found in their gas phase chemical behaviours, we assume acetaldehyde to behave similarly to its isomer. Moreover, we set in UCL\_CHEM the same initial elemental abundances as in the MONACO code (see Table 1).
\begin{figure*}
 \begin{center}
\includegraphics[width=130mm]{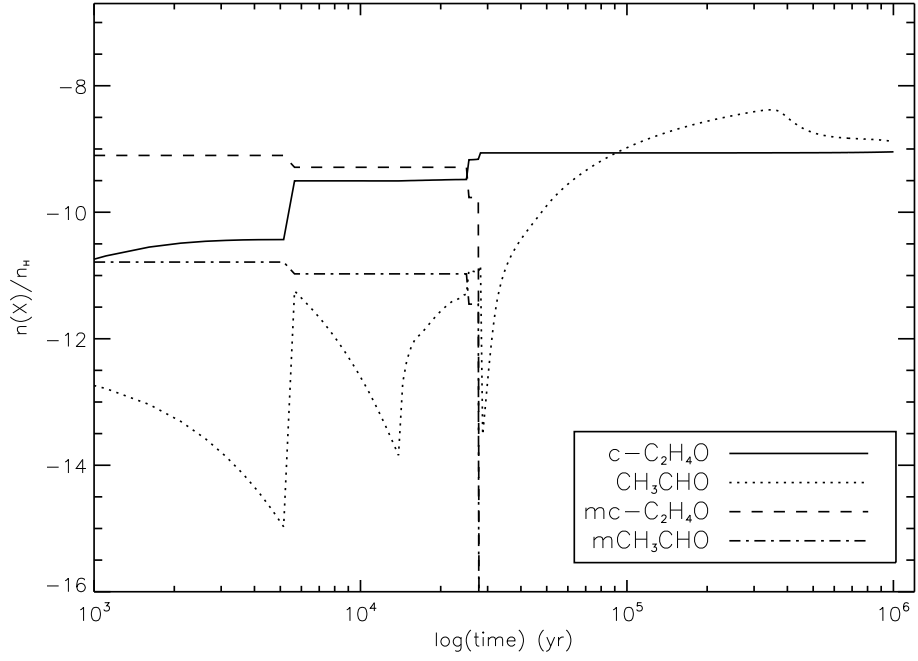}
\caption{Fractional abundances as a function of time (yrs) for ethylene oxide and acetaldehyde, both in the gas phase and on the grain-surface obtained as output from UCL\_CHEM model for the warm-up phase during the hot core formation, which starts once the collapse of the core is terminated and molecules are frozen onto the grain-surface. The designation $m$ refers to the grain mantle.}
 \end{center}
\end{figure*}

The output for ethylene oxide and acetaldehyde obtained after we updated the UCL\_CHEM network is plotted in Fig. 4. The plot shows the fractional abundances  of gaseous (solid line) and solid (dashed line) ethylene oxide as well as gaseous (dotted line) and solid (dashed-dotted line) acetaldehyde, during the warm-up phase leading to a hot core.
For both isomers we clearly observe the existence of a sharp desorption peak during which the gas phase abundance first increases sharply and the solid-state abundance decreases. The sharp decrease in the solid state abundances of both molecules at 6$\times$10$^{3}$ years occurs at even lower temperatures ($\sim$43-45 K) than those predicted by running the MONACO model.

Volcanic and co-desorption effects can also be highlighted at late stages (just before 3$\times$10$^{4}$ years) during the protostellar formation and they arise at $\sim$90 K and $\sim$100 K, respectively. 
Moreover, comparing the results obtained from the two codes, we find general agreement in predicting the sub-thermal sublimation of the two isomers.
Another important consideration concerns the ratio between the acetaldehyde-to-ethylene oxide abundances in the gas phase at the end of the warm-up phase ($\sim$1.2$\times$10$^{6}$ yrs). From UCL\_CHEM, we found a value of $\sim$ 1.5, which fits well within the interval of values reported in the literature \citep{Nummelin98, Ikeda01}.

We therefore compare the new fractional abundances obtained by running UCL\_CHEM with those from MONACO. The results are summarised in Table 5. Note that the UCL\_CHEM value for ethylene oxide abundance has increased from the value in Table 4 by a factor of 9, due principally to reactions involving the conversion of acetaldehyde, newly added to the model, to ethylene oxide.

\begin{table}
 \begin{center}
  \caption{Comparison between the fractional abundances of the two isomers obtained from MONACO and UCL\_CHEM models (at 200 K).}
  \begin{tabular}{@{}ccc}
  \hline
 & c-C$_{2}$H$_{4}$O & CH$_{3}$CHO\\
   \hline
MONACO & 2$\times$10$^{-9}$ & 2$\times$10$^{-9}$\\
UCL\_CHEM & 9$\times$10$^{-10}$ & 1$\times$10$^{-9}$ \\
   \hline
  \end{tabular}
 \end{center}
\end{table} 

 In addition to ethylene oxide and acetaldehyde, our outputs from MONACO models contain predicted abundances of other hot core tracers. In Fig. 5, the fractional abundances of a number of these species have been plotted as a function of time during the collapse and the warm-up phases of hot core formation.  
\begin{figure*}
 \begin{center}
\includegraphics[width=90mm, angle=450]{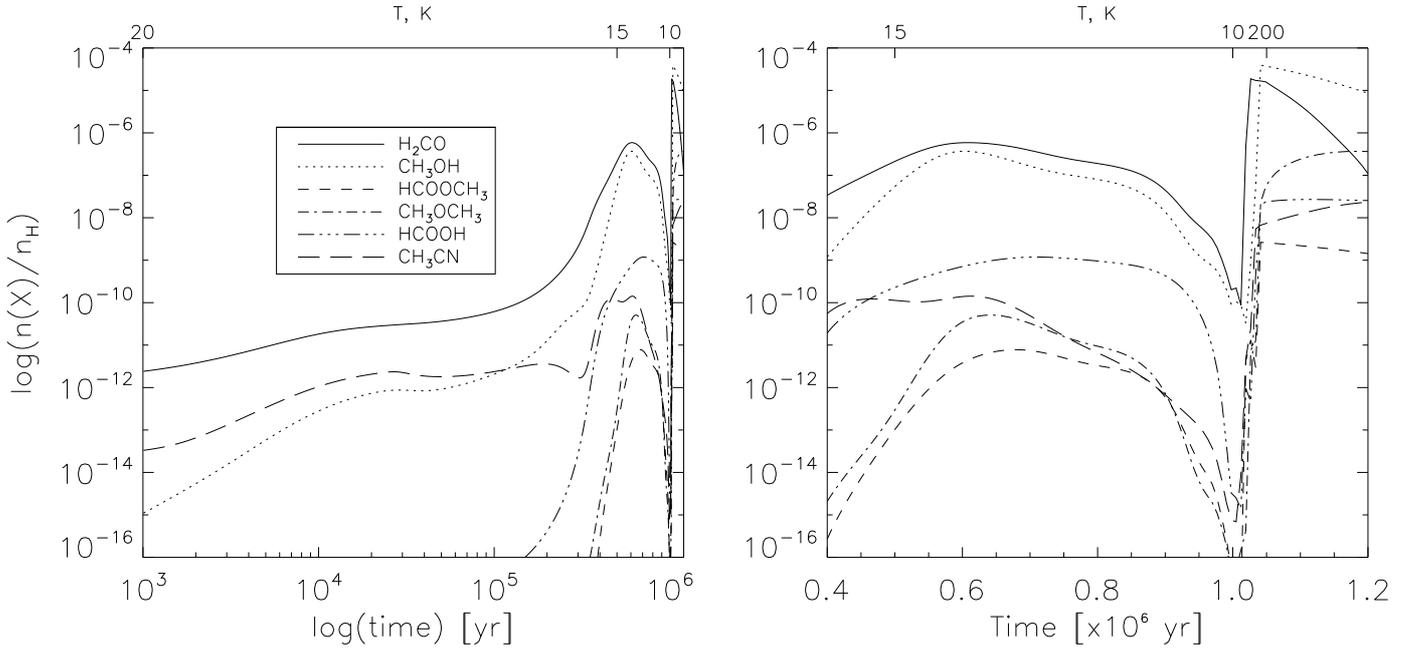}
\caption{Fractional abundances (with respect to the total number of hydrogen nuclei) as a function of time (yr) for various hot core tracers. The panel on the right expands the final portion of the panel on the left.}
 \end{center}
\end{figure*}
We have also compared our theoretical results and data from recent observations towards SgrB2 sources \citep{Belloche13}. To allow this comparison, we converted our theoretical fractional abundances at 1.2$\times$10$^{6}$ yrs into column densities (in cm$^{-2}$) for each species by applying Equation (3).
The calculated column densities together with those reported in the literature are listed in Table 6. 
\begin{table}
 \begin{center}
  \caption{Comparison between theoretical and observed column densities (in cm$^{-2}$) for selected species.}
  \begin{tabular}{@{}ccc}
  \hline
 & Modelling & Observations\\
   \hline
c-C$_{2}$H$_{4}$O & 2$\times$10$^{17}$ & 4$\times$10$^{16}$\\
CH$_{3}$CHO & 2$\times$10$^{17}$ & 1$\times$10$^{17}$\\
CH$_{3}$OH & 3$\times$10$^{21}$ & 1$\times$10$^{19}$\\
CH$_{3}$OCH$_{3}$ & 4$\times$10$^{19}$ & 9$\times$10$^{17}$\\
H$_{2}$CO & 1$\times$10$^{21}$ & 5$\times$10$^{17}$\\
HCOOH & 2$\times$10$^{18}$ & 1$\times$10$^{16}$\\
HCOOCH$_{3}$ & 3$\times$10$^{17}$ & 1$\times$10$^{16}$\\
CH$_{3}$CN & 5$\times$10$^{17}$ & 7$\times$10$^{17}$\\
   \hline
  \end{tabular}
 \end{center}
\centering
\tiny Observations are taken from \citet{Belloche13}. Note that the theoretical column densities are calculated based on the fractional abundances of molecules obtained as output from MONACO code at 1.2$\times$10$^{6}$ yrs.
\end{table} 

As we can see from the entries in Table 6, there is reasonable agreement; i.e., within one to two orders of magnitude, between our theoretical column densities and those taken from the study by \citet{Belloche13}. The only exception is H$_{2}$CO, which is overestimated by more than three orders of magnitude. Moreover, note that, although we are able to reproduce the final abundance of CH$_{3}$OH, this species shows a very high (and probably not realistic) abundance during the cold phase due to the high efficiency of reactive desorption (see Fig. 5, left panel).

\section{Conclusions}
We used the MONACO code, as implemented by \citet{Vasyunin13}, to model  the formation and destruction of ethylene oxide and acetylene for a typical hot core. In particular, we extended the
 MONACO model to include recent experimental results \citep{Ward11} on ethylene oxide formation on icy mantle analogues by the association of oxygen atoms and ethylene.
For comparison, we considered a previous study by \citet{Occhiogrosso12} focused on the formation of solid ethylene oxide and we inserted into UCL\_CHEM code, a reaction network for the formation and the loss pathways of acetaldehyde. We then ran a model with the same physical parameters as in the simulation performed using the MONACO model. Our main conclusions from this computational analysis are:

   \begin{enumerate}
      \item By employing the two different codes, ethylene oxide and acetaldehyde show an increase in their gaseous abundances at low temperatures during the warm-up phase of the star formation process. In particular, both codes for ethylene oxide display a plateau after they reach their peak abundances, while, in both cases, the amount of acetaldehyde slightly decreases (after its peak value) due to the reaction with H$_{3}$O$^{+}$ to form C$_{2}$H$_{5}$O$^{+}$ and H$_{2}$O (see reaction 11).
      \item The theoretical evidence of early desorption at relatively low temperatures suggests that ethylene oxide might exist at an observable level in the outer and cooler regions of hot cores, where its isomer has already been detected. 
      \item Both models show an efficient formation of ethylene oxide and acetaldehyde on grains already at early stages during the collapse phase. After $\sim$ 10$^{6}$ years the importance of the surface reaction for ethylene oxide formation becomes comparable to its production in the gas phase thanks to the loop reactions in which acetaldehyde is destroyed to form C$_{2}$H$_{5}$O$^{+}$ (which is in turn one of the reactants to produce gaseous ethylene oxide, as shown in reaction 8). 
      \item The theoretical ratios between the fractional abundances of the two isomers evaluated with the UCL\_CHEM and MONACO models fit well within the range of values reported in the literature for the case of a hot core source \citep{Nummelin98, Ikeda01}. This factor may be important to quantify the amount of ethylene oxide towards cold regions based on the acetaldehyde abundance in these astronomical environments. Moreover, the calculated column densities of both isomers (at 200 K and 1.2$\times$10$^{6}$ yrs) perfectly match those from recent observations towards SgrB2 sources \citep{Belloche13}.
      \item A final important consideration regards the theoretical treatment of the experimental results. Despite the fact that the same laboratory data \citep{Ward11} have been used in the present study as well as in \citet{Occhiogrosso12}, different assumptions were made in the two cases in converting experimental data for use in the interstellar environment. The consistency between the two codes in reproducing the ISM chemistry of ethylene oxide and acetaldehyde may be a confirmation of the reliability of the scientific justification asserted in both studies.
   \end{enumerate}

\begin{acknowledgements}
The research leading to these results has received funding from the [European Community's] Seventh Framework Program [FP7/2007-2013] under grant agreement n$^{\circ}$ 238258.    
E. H. acknowledges the support of the National Science Foundation (US) for his astrochemistry program, and  support from the NASA Exobiology and Evolutionary Biology program through a subcontract from Rensselaer Polytechnic Institute.  
\end{acknowledgements}

\bibliographystyle{bibtex/aa}
\bibliography{paper4} 

\label{lastpage}
\end{document}